\documentclass[conference]{IEEEtran}
\IEEEoverridecommandlockouts
\usepackage{cite}
\usepackage{amsmath,amssymb,amsfonts}
\usepackage{balance}

\newtheorem{theorem}{Theorem}
\newtheorem{lemma}[theorem]{Lemma}
\newtheorem{define}[theorem]{Definition}
\newtheorem{assume}[theorem]{Assumption}
\newtheorem{remark}[theorem]{Remark}

\usepackage{algorithmic}
\usepackage{graphicx}
\usepackage{subfig}

\usepackage{textcomp}
\usepackage{xcolor}
\def\BibTeX{{\rm B\kern-.05em{\sc i\kern-.025em b}\kern-.08em
    T\kern-.1667em\lower.7ex\hbox{E}\kern-.125emX}}

\usepackage{tikz}
\usepackage{tikz,tkz-euclide}
\usetikzlibrary{calc}
\usetikzlibrary{decorations.markings}
\tikzset{>=latex} 
\usetikzlibrary{backgrounds}
\usetikzlibrary{positioning}
\usepackage{subfig}
\usetikzlibrary{shapes.geometric, arrows}

\begin{document}
 
\title{Bearing-Based Network Localization Under Randomized Gossip Protocol%*\\
\author{Nhat-Minh Le-Phan, Minh Hoang Trinh$^*$, Phuoc Doan Nguyen
\thanks{N.-M. Le-Phan and P. D. Nguyen are with Department of Automation Engineering, School of Electrical and Electronic Engineering, Hanoi University of Science and Technology, Hanoi, Vietnam. N.-M. Le-Phan is also with Viettel Aerospace Institute, Hanoi, Vietnam. E-mails: \texttt{minh.lpn221013m@sis.hust.edu.vn; phuoc.nguyendoan@hust.edu.vn}. M. H. Trinh is with FPT University, Quy Nhon AI Campus, Binh Dinh, Vietnam. Email: \texttt{minhtrinh@ieee.org}.
}
\thanks{$^*$Corresponding author.}
}
}
%\author{Nhat-Minh Le-Phan, Minh Hoang Trinh$^*$, Phuoc Doan Nguyen}

\maketitle

\begin{abstract}
In this paper, we consider a randomized gossip algorithm for the bearing-based network localization problem. Let each sensor node be able to obtain the bearing vectors and communicate its position estimates with several neighboring agents. Each update involves two agents, and the update sequence follows a stochastic process. Under the assumption that the network is infinitesimally bearing rigid and contains at least two beacon nodes, we show that when the updating step-size is properly selected, the proposed algorithm can successfully estimate the actual sensor nodes' positions with probability one. The randomized update provides a simple, distributed, and cost-effective method for localizing the network. The theoretical result is supported with a simulation of a 1089-node sensor network. 
\end{abstract}


\section{Introduction}
\label{sec:introduction}
With the revolution of the next-generation network in recent years, the topic of network localization has been studied more widely by researchers due to its role in both network operations and many application tasks. For example, in a sensor network, the sensor nodes must be aware of their precise locations in order to route packets via geometric routing, and record and detect events \cite{Aspnes2006}. GPS could be a solution, but the cost of GPS devices and the non-availability of GPS signals in restricted environments prevent their use in large-scale sensor networks. Thus, network localization algorithms, which estimate the locations of sensors with initially unknown location information by using knowledge of the absolute positions of a few sensors (beacons) and inter-sensor measurements such as distance and bearing measurements are preferred \cite{MAO20072529}. 

In this paper, we focus specifically on the case where sensors are able to obtain bearing measurements and cannot measure distances. Compared to distance-based and position-based network localization, bearing sensing capability is a minimal requirement of the agent. In the real world, bearing measurements can be obtained by an on-board camera, which is passive and transmits no signal \cite{Ye2017bearing}. Due to its advantages, bearing-based network localization has
attracted extensive research attention recently, see for example, \cite{Bishop2009bearing,zhu2014, Zhong2014} on application to networks in two-dimensional spaces; \cite{ZHAO2016334, Li2020tcns} on works with three and higher dimensional spaces; \cite{Cao2021icsmc} on dealing with the case where the common global reference frame does not hold. It is noted that the sensor network can be considered as a matrix-weighted graph, where, the connections between sensors/agents are represented by matrices relating to bearing vectors.

Gossip algorithms \cite{boyd2006,ishii2010}, in which the communications between sensors are randomly selected for each discrete instant, have received a lot of attention in several areas such as distributed computation, network optimization, and  wireless systems. The main advantages of this algorithm class are low-cost communication requirements (each sensor communicates with one neighbor at a time) and robustness with communication link failure \cite{STRAKOVA2013IOP}. A number of researchers have investigated several variations of the classic gossip algorithm., see for examples, \cite{5625615,TIT2010,1238221} on geographic gossip; \cite{7581065,4787122} on broadcast gossip; \cite{8946047} on reduce the probability of selecting duplicate nodes, \cite{Liu2013analysis} on accelerating the convergence speed, \dots In \cite{arxivminh}, authors proposed a gossip-based matrix-weighted consensus algorithm, dealing with the case that the weights between agents are represented by positive semi-definite matrices. 

The main contribution of this paper is proposing a gossip-based network localization algorithm for arbitrary dimension space using only bearing measurements and exchanged position estimates. Several conditions for the convergence of the proposed algorithm and an estimate of the convergence time are also derived. Although this paper only asserts the effectiveness of the algorithm for leader-follower network, a similar analysis holds for undirected networks.

The remainder of this paper is organized as follows. In Section \ref{sec:preliminaries}, we introduce the preliminaries and problem formulation. Our main analysis are stated in Section \ref{sec: mainresults}. The simulation results are provided in Section \ref{simulation}. Finally, we will draw our conclusions and provide directions for future research in Section \ref{conclusion}.

\section{Preliminaries and problem formulation} \label{sec:preliminaries}
\subsection{Expected Matrix-weighted Graph}
A matrix-weighted graph \cite{HMT2018} is denoted by $\mathcal{G}=(V,E,A)$, where, $V=\left\{ 1,2,...,n\right\}$ is the vertex set (agents), $E\subseteq V\times V$ is the edge set, and 
$A=\{ \mathbf{A}_{ij} \in \mathbb{R}^{d \times d}|~(i,j) \in E\}$ denotes the set of matrix weights.\footnote{Note that $d \geq 1$ is the dimension of each agent's state vector. When $d = 1$, $\mathcal{G}$ reduces to a scalar graph.} The interactions between any two agents in $\mathcal{G}$ are captured by the corresponding matrix weights. If $(i,j)\in E$, there is a symmetric positive definite/positive semi-definite matrix weight $\mathbf{A}_{ij}=\mathbf{A}_{ij}^\top\geq 0$; and if $i$ and $j$ are disconnected, then  $\mathbf{A}_{ij} = \mathbf{0}_{d\times d}$. %Note that  $(i,j)$ and $(j,i)$ may correspond to two distinct nonnegative definite matrix weights $\mathbf{A}_{ij}$ and $\mathbf{A}_{ji}$.\footnote{In this paper, a matrix $\mathbf{A}_{ij} \in \mathbb{R}^{d\times d}$ is positive semidefinite (positive definite) if and only if $\mathbf{x}^\top\mathbf{A}_{ij}\mathbf{x} \geq 0,~\forall \mathbf{0}_d \neq \mathbf{x} \in \mathbb{R}^d$ (resp., $\mathbf{x}^\top\mathbf{A}_{ij}\mathbf{x} > 0,~\forall \mathbf{0}_d \neq \mathbf{x} \in \mathbb{R}^d$).} 

Let $\mathcal{G}^{\rm M} = (V,E^{\rm M}, {A}^{\rm M})$ be the expected matrix weighted graph corresponding to $G$, then $\mathcal{G}^{\rm M}$ is undirected and has the vertex set $V$, the edge set $E^{\rm M} = \{(i,j) |~\exists (i,j) \in E,~i, j \in V\}$, and the set ${A}^{\rm M}$ of expected matrix weights $\mathbf{M}_{ij}= \mathbf{M}_{ij}^\top = \mathbf{M}_{ji} =\frac{1}{n}(\mathbf{A}_{ij} {\rm P}_{ij}+\mathbf{A}_{ji} {\rm P}_{ji})$ between $i$ and $j$ (${\rm P}_{ij}, {\rm P}_{ji} \in [0,1]$  are probabilities). We call an edge $(i,j)$ positive definite (resp., positive semi-definite) if the associated expected weight $\mathbf{M}_{ij}$ is positive definite (resp., positive semi-definite). The \emph{expected degree matrix} is defined as $\mathbf{D}^{\text{M}}= \text{blkdiag}(\mathbf{D}^{\text{M}}_{1},\ldots,\mathbf{D}^{\text{M}}_{n})$, where $\mathbf{D}^{\text{M}}_{i}= \sum_{j \in V}\mathbf{M}_{ij}$. Then, $\mathbf{L}^{\text{M}} = \mathbf{D}^{\text{M}}-\mathbf{A}^{\text{M}} \in \mathbb{R}^{nd \times nd}$ is the \emph{expected matrix-weighted Laplacian} of $\mathcal{G}$.\footnote{The matrix can also be referred to as the expected bearing Laplacian to be consistent with the terminology in \cite{zhaoTAC}.} 
\begin{lemma}\label{nullspaceL}
\emph{\cite{HMT2018}}
The expected Laplacian matrix $\mathbf{L}^{\rm M}$ is symmetric and positive semi-definite, and its null space is given as:
  ${\rm null}(\mathbf{L}^{\rm M}) = {\rm span}\{ {\rm range}(\mathbf{1}_n \otimes \mathbf{I}_d), \{ \mathbf{v}=[v_1^\top,\dots, v_n^\top]^\top \in \mathbb{R}^{nd}|~(v_j-v_j)\in {\rm null}(\mathbf{M}_{ij}), \forall (i,j) \in~E \} \}$.
\end{lemma}

\begin{lemma}
\emph{(Markov inequality)}\cite{prob} If a random variable $X$ can only take non-negative values, then 
   $${\rm P}(X>a) \leq \frac{{{\rm E}}[X]}{a}, \ \  \forall a >0,$$
where ${\rm E}[X]$ is the expectation of $X$.
\end{lemma}

\subsection{Bearing Rigidity Theory}
The bearing rigidity theory plays an important role in the analysis of bearing-based network localization problems. In this section, we will go through a few key concepts and results from the bearing rigidity theory \cite{bearingzhao}. 

Consider a sensor network of $n$ nodes (or agents) in $\mathbb{R}^d$ $(n \geq 2, d \geq 2)$. Each agent $i \in \left\{ 1,2,...,n\right\}$ has an absolute position $p_i \in \mathbb{R}^d$ (which needs to be estimated). Suppose that $p_i \neq p_j$, the \emph{bearing vector} between two agents $i$ and $j$ is defined as \cite{bearingzhao}
\begin{equation}
    g_{ij}=\frac{p_i-p_j}{||p_i-p_j||}.
\end{equation}
It can be checked that $\|g_{ij}\|=1$ as $g_{ij}$ is a unit vector. 

Let the sensor network have a underlying matrix-weighted graph $\mathcal{G}=(V,E,A)$, where $V=\left\{ 1,2,...,n\right\}$ is the vertex set (agents), $E\subseteq V\times V$ is the edge set ($E=\{e_1,\ldots,e_m\}=\{\ldots,e_{ij},\ldots\}$), and the matrix weights in $A$ are orthogonal projection matrices\footnote{An orthogonal projection corresponding to vector $x \in \mathbb{R}^d$ is transforms a vector $y \in \mathbb{R}^d$ into the closest point with $y$ that belongs to the orthogonal complement of $x$.}
\begin{equation}
    \mathbf{A}_{ij}=\mathbf{I}_d-{g_{ij}g_{ij}^{\top}}.%{||g_{ij}||^2}
\end{equation}
It is clear that $\mathbf{A}_{ij}=\mathbf{A}_{ji}=\mathbf{A}_{ij}^{\top}$ (symmetric). Furthermore, $\mathbf{A}_{ij}^2=\mathbf{A}_{ij}\geq 0$ (idempotent and positive semidefinite), $\text{Null}(\mathbf{A}_{ij}) = \text{span}(g_{ij})$ and
$\mathbf{A}_{ij}$ has one zero eigenvalue and $d-1$ unity eigenvalues \cite{bearingzhao}. 

A framework (or a network) is defined by $\mathcal{G}(p)$, where $\mathcal{G}$ is a matrix weighted graph, and $p=[p_1^{\top},p_2^{\top},...,p_n^{\top}]^{\top} \in \mathbb{R}^{dn}$ is a configuration in the $d$-dimensional space. The \emph{bearing function} of a \emph{framework} $\mathcal{G}(p)$ is defined as \[ F_B(p) = [\ldots,g_{ij}^\top,\ldots]^\top = [g_1^{\top},g_2^{\top},...,g_m^{\top}]^{\top}\in \mathbb{R}^{dm},\] where the bearing vector $g_k=g_{ij}$ corresponds to the $k$-th edge $e_k=(i,j)$ in $E$. In other words, the bearing function contains all bearing vectors that constrain the locations of nodes in the network. The \emph{bearing rigidity matrix} is defined as the Jacobian of the bearing function \cite{bearingzhao}
\begin{equation}
    R_B(p) = \frac{\partial F_B(p)}{\partial p} \in \mathbb{R}^{dm \times dn}.
\end{equation}

The augmented bearing rigidity matrix can be expressed as \[R_B(p)=\text{diag}\left(\frac{\mathbf{A}_{k}}{\|p_i-p_j\|}\right)(H\otimes \mathbf{I}_d),\] 
where $H \in \mathbb{R}^{m \times n}$ is the incidence 
matrix corresponding to an arbitrary ordering and orientation of the edges in $E$.

We next introduce the definition of infinitesimally bearing rigid framework, which can be found in \cite{bearingzhao}. 

\begin{define}
 A framework $\mathcal{G}(p)$ is infinitesimally bearing rigid if and only if the motion preserves the bearing function of the framework are trivial, i.e., translation and scaling.  
\end{define}
\begin{theorem}\label{Rb}
\emph{\cite{bearingzhao}} The infinitesimally bearing rigidity of $\mathcal{G}(p)$ is equivalent to
\begin{enumerate}
    \item $\text{rank}(R_B)= dn-d-1$,
    \item $\text{Null}(R_B)=\text{span}\left\{\mathbf{1}_n\otimes \mathbf{I}_d,p\right\}$.
\end{enumerate}
\end{theorem}

\begin{figure}
\centering
\subfloat[]{%{\resizebox{3cm}{!}
\begin{tikzpicture}[
roundnode/.style={circle, draw=black, thick, minimum size=2mm,inner sep= 0.25mm},
squarednode/.style={rectangle, draw=red!60, fill=red!5, very thick, minimum size=5mm},
]
    \node[roundnode]   (v1)   at   (1,1) {$1$};%   {\footnotesize $v_1$};
    \node[roundnode]   (v2)   at   (0,0) {$2$};%   {\footnotesize $v_1$};
    \node[roundnode]   (v3)   at   (0,1) {$3$};%   {\footnotesize $v_1$};
    \node[roundnode]   (v4)   at   (-1,0) {$4$};%   {\footnotesize $v_1$};
    \draw[-,thick] (v1)--(v2)--(v4)--(v3)--(v2);
    \draw[-,thick] (v1)--(v3);
\end{tikzpicture} \label{rigid example}
}
\qquad
\subfloat[]{ %{\resizebox{3cm}{!}
\begin{tikzpicture}[
roundnode/.style={circle, draw=black, thick, minimum size=2mm,inner sep= 0.25mm},
squarednode/.style={rectangle, draw=red!60, fill=red!5, very thick, minimum size=5mm},
]
    \node[roundnode]   (v1)   at   (1,1) {$1$};%   {\footnotesize $v_1$};
    \node[roundnode]   (v2)   at   (0,0) {$2$};%   {\footnotesize $v_1$};
    \node[roundnode]   (v3)   at   (0,1) {$3$};%   {\footnotesize $v_1$};
    \node[roundnode]   (v4)   at   (-1,0) {$4$};%   {\footnotesize $v_1$};
    \draw[-,thick] (v1)--(v2)--(v4)--(v3);
    \draw[-,thick] (v1)--(v3);
\end{tikzpicture}
\label{norigid example}}
%    \subfloat[Non-infinitesimally bearing rigid framework.]{\includegraphics[width=0.45\columnwidth]{norigid.png}\label{norigid example}}\\
    \caption{Examples of infinitesimally/non-infinitesimally bearing rigid frameworks in two-dimensional space.}
\end{figure}
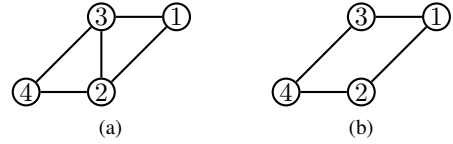

An example of infinitesimally/non-infinitesimally bearing rigid frameworks in the two-dimensional space is depicted in Fig.~\ref{rigid example}-\ref{norigid example}. The following assumption is employed in this paper.
\begin{assume}\label{bearing rigid}
    The network $\mathcal{G}(p)$ is infinitesimally bearing rigid.
\end{assume}

The bearing-based network localization problem can be stated as follows.

\textbf{Problem.} \emph{Let Assumption \ref{bearing rigid} hold and suppose that there exist at least $2\leq n_a< n$ beacon nodes which know their absolute positions. The initial position estimation of the system is $\hat{p}(0)$. Design the update law for each agent $u_i(k)=\hat{p}(k+1)-\hat{p}(k)$ $\forall i \in V$ based on the relative estimates $\left\{\hat{p}_i(k)-\hat{p}_j(k)\right\}$ and the constant bearing measurements $\left\{g_{ij}\right\}$ such that $\hat{p}(k)\rightarrow p$ as $k\rightarrow \infty$ for all $i\in V$.} 
%\subsection{Gossip protocol/Decide to omitt or not}
 %In this section, we describe the basic protocol that will be used for the rest of this article, which can be found in \cite{ishii2010}. Consider the multiagent system from the previous section. The random manner is specified by a random process $\gamma(k)\in V$ where $k\in \mathbb{Z}_{+}$ is called time slot. At time slot $k$, $\gamma(k) = i$ indicates that agent $i$ wakes up, then it will choose another neighbor $j$ with a probability ${\rm P}_{ij}$ to communicate, and both agents update their state values. We assume $\gamma(k)$ to be i.i.d., and its probability distribution is uniform
%\begin{equation}\label{manner}
%{\rm P}(\gamma(k)=i)=\frac{1}{n}, \forall k\in \mathbb{Z}_{+}.
%\end{equation}
%By providing each agent with a separate clock that "ticks" (wakes the agent up) at the time of identical stochastic processes, this protocol can be distributedly implemented. %In this paper, we do not restrict to any particular stochastic process for the purpose of simplicity. 
%We will use time slots as a new time axis in all of our results because they are the only instances in which each agent's value changes.

\section{MAIN RESULTS}\label{sec: mainresults}
In this section, we firstly present the randomized gossip algorithm for bearing-based network localization. Secondly, we specify sufficient conditions for the convergence in expectation of the algorithm's first- and second moments. Finally, a discussion on the convegence rate is also given. 
\subsection{Bearing-Based Network Localization Algorithm}

Consider a network consisting of $n$ sensors (agents) whose interconnections between agents $\mathbf{A}_{ij}\in\mathbb{R}^{d\times d}$ are defined in Section \ref{sec:introduction}. Suppose there are $n_a$ agents ($0\leq n_a \leq n$), known as beacons, can measure their own real positions. The rest $n_f=n-n_a$ agents are
called followers. (Note that the network cannot be localized without beacons). %Each follower $i\in V$ stores its own estimation vector $\overline{x}_i \in \mathbb{R}^d$. 
The randomized manner is specified by a random process $\gamma(k)\in V$ where $k\in \mathbb{Z}_{+}$ is called a time slot. At time slot $k$, $\gamma(k) = i$ (with probability $\frac{1}{n}$) indicates that agents $i$ wakes up, then it will choose another neighbor $j$ with a probability ${\rm P}_{ij}$ to communicate. If both the waken and chosen ones are beacons, then they just retain their values. If both of the waken and chosen ones are followers, they will update their values as an algorithm in (\ref{algorithm1}). If one of the two agents is a beacon and the other is a follower, only the follower can update its value. In summary, the updating law is designed as follows 
%When a follower $i$ wakes up at the $k^{\text{th}}$ time slot, it will contact a neighbor $j$ with a probability ${\rm P}_{ij}$ (The waken agent cannot choose itself, i.e., ${\rm P}_{ii}\neq0$) and both agents will update their current state vector as:
%
\begin{enumerate}
\item if $i$ and $j$ are beacons:
\begin{equation}\label{algorithm0}
    \begin{aligned}
    {p}_i(k+1)&={p}_i(k)=p_i,\\
    {p}_j(k+1)&={p}_j(k)=p_j.
    %W_{ij} &= \mathbf{I}_{dn\times dn}
    \end{aligned}
\end{equation}

\item if $i$ and $j$ are followers:
\begin{equation}\label{algorithm1}
    \begin{aligned}
    \hat{p}_i(k+1)&=\hat{p}_i(k)-\alpha \mathbf{A}_{ij}\big(\hat{p}_i(k)-\hat{p}_j(k)\big),\\
    \hat{p}_j(k+1)&=\hat{p}_j(k)-\alpha \mathbf{A}_{ji}\big(\hat{p}_j(k)-\hat{p}_i(k)\big).\\
    \end{aligned}
\end{equation}
\item if one of the partners is a beacon and the other is a follower (without loss of generality, assume $i$ is a follower)
\begin{equation}\label{algorithm2}
    \begin{aligned}
    \hat{p}_i(k+1)&=\hat{p}_i(k)-\alpha \mathbf{A}_{ij}\big(\hat{p}_i(k)-{p}_j(k)\big),\\
    {p}_j(k+1)&={p}_j(k)=p_j,
    \end{aligned}
\end{equation}
\end{enumerate}
where $\alpha > 0$ is updating step size and will be designed later for guaranteeing the convergence of the algorithm.

Without loss of generality, we denote the first $n_a$ agents as beacons ($V_a = \left\{1,2,...,n_a\right\}$) and the rest as followers ($V_f = \left\{n_a+1,n_a+2,...,n\right\}$). Denote $p_a=[p_1^{\top},p_2^{\top},...,p_{n_a}^{\top}]^{\top}$ and $p_f=[p_{n_a+1}^{\top},p_{n_a+2}^{\top},...,p_{n}^{\top}]^{\top}$.

%\begin{remark}
%Due to the fact that just followers can be waken up, $P_{ij}=0$  $\forall i \in V_a$.
%\end{remark}

\begin{assume}\label{probability}
For every $(i,j)\in E$ such that $g_{ij} \in F_B$, ${\rm P}_{ij}+{\rm P}_{ji}>0$.   
\end{assume}

It can be seen that the probability that two agents $i$ and $j$ communicate with each other is $\frac{1}{n}({\rm P}_{ij}+{\rm P}_{ji})$ (the probability that agent $i$ will wake up at $k^{\text{th}}$ time slot is $\frac{1}{n}$, and the probability that $j$ will be chosen by $i$ is ${\rm P}_{ij}$). The expected Laplacian matrix, which was defined in  Section~\ref{sec:preliminaries}, can be partitioned into the following form 
$$\mathbf{L}^{\rm M}(\mathcal{G})=\begin{bmatrix}
\mathbf{L}^{\rm M}_{aa} &\mathbf{L}^{\rm M}_{af} \\
\mathbf{L}^{\rm M}_{fa} &\mathbf{L}^{\rm M}_{ff} \\
\end{bmatrix},$$
where
\begin{equation}
 \begin{aligned}
 %\mathbf{L}^{\rm M}_{aa}&=0_{n_ad\times n_ad}\\
 [\mathbf{L}^{ \rm M} ]_{ij}&=-\frac{1}{n}(\mathbf{A}_{ij} {\rm P}_{ij}+\mathbf{A}_{ji} {\rm P}_{ji})=-\mathbf{M}_{ij}, \ \ i\neq j,\\
[\mathbf{L}^{\rm M}]_{ii}&=\frac{1}{n}\sum_{j\in \mathcal{N}_i}(\mathbf{A}_{ij} {\rm P}_{ij}+\mathbf{A}_{ji} {\rm P}_{ji})=\sum_{j\in \mathcal{N}_i}\mathbf{M}_{ij}.
 \end{aligned}
\end{equation}

\begin{remark}
%It is notable that $\mathbf{A}_{ij}=\mathbf{A}_{ji}=\mathbf{A}_{ij}^{\top}$. Thus, $\mathbf{M}_{ij}=\mathbf{M}_{ji}=\mathbf{M}_{ij}^{\top}$. Moreover, 
Assumption~\ref{probability} implies that $\mathbf{M}_{ij}$ is positive definite (resp., positive semi-definite) if and only if $\mathbf{A}_{ij}$ is positive definite (resp., positive semi-definite).
\end{remark}

Taking the expectation of \eqref{algorithm0}-\eqref{algorithm2}, the following equations could be obtained
\begin{equation}\label{sys}
    \begin{aligned}
        \hat{p}_a(k+1)&=\hat{p}_a(k)=p_a,\\
        \overline{\hat{p}}_f(k+1)&=(\mathbf{I}_{n_f}-\alpha \mathbf{L}^{\rm M}_{ff})\overline{\hat{p}}_f(k)-\alpha\mathbf{L}^{\rm M}_{fa}p_a,
    \end{aligned}
\end{equation}
where $\overline{\hat{p}}(k)$ is the expectation of $\hat{p}(k)$.

\begin{lemma}
\text{\emph{\cite{bearingzhao}}}
 Under Assumption \ref{bearing rigid}, the matrix $\mathbf{L}^{\rm M}_{ff}$ is positive definite if and only if $n_a \geq 2$.
\end{lemma}

\begin{lemma}\label{null}
    Under Assumption~\ref{bearing rigid}, the expected Laplacian is symmetric positive semi-definite. Moreover, it satisfies $\text{rank}(\mathbf{L}^{\rm M})=dn-d-1$ and $\text{Null}(\mathbf{L}^{\rm M})=\text{span}\left\{\mathbf{1}\otimes\mathbf{I}_d,p\right\}$
\end{lemma}
\begin{IEEEproof} $\mathbf{L}^{\rm M}$ can be written as \cite{bearingzhao} $$\mathbf{L}^{\rm M}=(H\otimes \mathbf{I}_d)^{\top}\text{diag}(\mathbf{M}_k)(H\otimes \mathbf{I}_d)$$ where $k=1,2,...,m$. In addition, \[\mathbf{M}_k=\frac{1}{n}({\rm P}_{ij}+{\rm P}_{ji})\mathbf{A}_{k} =\mathbf{A}_{k}^{\top}\frac{1}{n}({\rm P}_{ij}+{\rm P}_{ji})\mathbf{A}_{k}.\] Thus, we represent $\mathbf{L}^{\rm M}$ as
$$\mathbf{L}^{\rm M}=\underbrace{(H\otimes \mathbf{I}_d)^{\top}\mathbf{A}_{k}^{\top}}_{:=\bar{R}_B^{\top}} \Big(\text{diag}\Big(\frac{{\rm P}_{ij}+{\rm P}_{ji}}{n}\Big)\otimes I_d \Big)\underbrace{\mathbf{A}_{k}(H\otimes \mathbf{I}_d)}_{:=\bar{R}_B}$$
Under Assumption~\ref{probability}, $\text{diag}\big(\frac{1}{n}({\rm P}_{ij}+{\rm P}_{ji})\big)\otimes I_d$ is positive definite. It is easy to prove that the expected Laplacian matrix and bearing rigidity matrix have the same rank and null space. Thus, Theorem~\ref{Rb} completes our proof.\end{IEEEproof}

\begin{lemma}
    Under Assumption~\ref{bearing rigid}, if network has at least two beacons, ${{p}}_f$ and $p_a$ satisfy
\begin{equation}
    {{p}}_f=-(\mathbf{L}_{ff}^{\rm M})^{-1}\mathbf{L}^{\rm M}_{fa}p_a.
\end{equation}
\end{lemma}
\begin{IEEEproof} From Lemma~\ref{null}, we have
$$\mathbf{L}^{\rm M}(\mathcal{G})p=\begin{bmatrix}
\mathbf{L}^{\rm M}_{aa} &\mathbf{L}^{\rm M}_{af} \\
\mathbf{L}^{\rm M}_{fa} &\mathbf{L}^{\rm M}_{ff} \\
\end{bmatrix}
\begin{bmatrix}
p_a \\
p_f\\
\end{bmatrix}
=0_{dn}.$$
Thus, $\mathbf{L}^{\rm M}_{fa}p_a+\mathbf{L}^{\rm M}_{ff}p_f=0_{dn_f}$. Since $\mathbf{L}^{\rm M}_{ff}$ is invertible, 
the following yields 
 ${{p}}_f=-(\mathbf{L}_{ff}^{\rm M})^{-1}\mathbf{L}^{\rm M}_{fa}p_a$. \end{IEEEproof}
 
\subsection{Convergence in Expectation}

\begin{lemma}\label{stepsize}
    Let the step size for each agent satisfy $\alpha~<~\frac{2}{\lambda_{\max}(\mathbf{L}^{\rm M}_{ff})}$, the eigenspectrum of the matrix $(\mathbf{I}_{n_f}-\alpha \mathbf{L}^{\rm M}_{ff})$ lies entirely in the interval $(-1,1)$, i.e., for $k \in \mathbb{N}$,   \[\lim_{k \to \infty}(\mathbf{I}_{n_f}-\alpha \mathbf{L}^{\rm M}_{ff})^k = 0_{df \times df}.\]
\end{lemma}
\begin{IEEEproof} Noting that the matrix $\alpha \mathbf{L}^{\rm M}_{ff}$ has already been proven to be symmetric and positive definite. It could be easily obtained that by choosing the stepsize $\alpha$ satisfying Lemma~\ref{stepsize}, $0<\lambda(\alpha \mathbf{L}^{\rm M}_{ff}))<2$. Thus, $-1<\lambda(\mathbf{I}_{n_f}-\alpha \mathbf{L}^{\rm M}_{ff}))<1$.\end{IEEEproof}

 \begin{theorem}\label{steady}
     Under Assumption~\ref{bearing rigid}, $n_a\geq 2$ and stepsize $\alpha$ is chosen as Lemma~\ref{stepsize}, the estimate configuration ${\hat{p}}(k)$ in system \eqref{algorithm0}--\eqref{algorithm2} converges in expectation to the actual configuration $p$ as $k \to \infty$.  
 \end{theorem} 
\begin{IEEEproof} 
Let $\overline{\tilde{p}}_f(k) = \overline{\hat{p}}_f(k) - p_f $, we have
\begin{equation}
\label{eq:est_config}
\begin{aligned}
\overline{\tilde{p}}_f(k+1) &=(\mathbf{I}_{d n_f}-\alpha \mathbf{L}^{\rm M}_{ff})\overline{\hat{p}}_f(k)-\alpha\mathbf{L}^{\rm M}_{fa}p_a + p_f,\\
&=(\mathbf{I}_{d n_f}-\alpha \mathbf{L}^{\rm M}_{ff})\overline{\hat{p}}_f(k) - (\mathbf{I}_{d n_f}-\alpha \mathbf{L}^{\rm M}_{ff})p_f\\
&=(\mathbf{I}_{d n_f}-\alpha \mathbf{L}^{\rm M}_{ff})\overline{\tilde{p}}_f(k),
\end{aligned}
\end{equation}
Lemma~\ref{stepsize} implies that (\ref{eq:est_config}) is exponentially stable, or $\lim_{k \to \infty}\overline{\tilde{p}}_f(k) = 0_{d n_f}$. Thus, it follows that
\[\lim_{k \to \infty}\overline{\hat{p}}_f(k) = \overline{\hat{p}}_f(\infty)=p_f.\] \end{IEEEproof}

%\subsection{Convergence Rate} 
\subsection{Convergence of Second Moment}
{\scriptsize 
\begin{figure*}
    \begin{equation} \label{eq:13}
W_{ij}=\begin{bmatrix}
    &\mathbf{I}_d        &\cdots     &0_{d\times d}             &\cdots    &0_{d\times d} &\cdots &0_{d\times d} \\
    &\vdots     &\ddots     &\vdots          &\ddots    &\vdots &\ddots &\vdots \\
    &0_{d\times d}        &\cdots  &\mathbf{I}_d-\alpha \mathbf{A}_{ij} &\cdots    &\alpha \mathbf{A}_{ij} &\cdots &0_{d\times d}  \\
    &\vdots     &\ddots     &\vdots          &\ddots    &\vdots &\ddots &\vdots \\
    &0_{d\times d}        &\cdots  &\alpha \mathbf{A}_{ji} &\cdots    &\mathbf{I}_d-\alpha\mathbf{A}_{ji} &\cdots &0_{d\times d}  \\
    &\vdots     &\ddots     &\vdots          &\ddots    &\vdots &\ddots &\vdots \\
    &0_{d\times d}        &\cdots     &0_{d\times d}             &\cdots    &0_{d\times d} &\cdots &\mathbf{I}_d
\end{bmatrix}.
\end{equation}
\end{figure*}}
Denote $\Tilde{p}_i(k)=\hat{p}_i(k)-p_i$ $\forall i \in V_f$ and $\Tilde{p}_f=[\Tilde{p}_{n_a+1}^{\top},\Tilde{p}_{n_a+2}^{\top},...,\Tilde{p}_{n}^{\top}]^{\top}$. We subtract both sides of (\ref{algorithm1}) and (\ref{algorithm2}) by $p_f$. In addition, due to the fact that $p_i-p_j \in \text{null} (\mathbf{A}_{ij})$, a quantity $\alpha\mathbf{A}_{ij}(p_i-p_j)$ is added to the right-hand side of every follower's equation of \eqref{algorithm1}) and \eqref{algorithm2}. Thus, we can rewrite \eqref{algorithm0}--\eqref{algorithm2} as
\begin{equation}
\Tilde{p}_f(k+1)=W_{ij}\Tilde{p}_f(k)
\end{equation}
where
\begin{enumerate}
\item if $i$ and $j$ are beacons:
\begin{equation}\label{algorithm01}
    \begin{aligned}
%    {p}_i(k+1)&={p}_i(k)\\
%    {p}_j(k+1)&={p}_j(k)\\
    W_{ij} &= \mathbf{I}_{dn_f\times dn_f}.
    \end{aligned}
\end{equation}

\item if $i$ and $j$ are followers, the updating matrix $W_{ij}$ is as given in \eqref{eq:13}, where $0_{d\times d}$ denotes the $d\times d$ zero matrix, $\mathbf{I}_d-\alpha \mathbf{A}_{ij}$ is the block entry of matrix $W_{ij}$ in the $({i(d-1)+1:id})^{\text{th}}$ rows
and $({i(d-1)+1:id})^{\text{th}}$ columns. Block $\mathbf{I}_d-~\alpha \mathbf{A}_{ij}$ is in the $({j(d-1)+1:jd})^{\text{th}}$ rows
and $({j(d-1)+1:jd})^{\text{th}}$ columns of $W_{ij}$. 
\item if one agent $i$ is a follower and the other agent $j$ is a beacon, we have the updating matrix 
\begin{equation}\label{algorithm21}
\begin{aligned}
W_{ij} = \text{blkdiag}(\mathbf{I}_d,\dots,\mathbf{I}_d-\alpha \mathbf{A}_{ij},\dots,\mathbf{I}_d).
\end{aligned}
\end{equation}
\end{enumerate}
It can be seen that for all three scenarios, $W_{ij}$ is symmetric due to the symmetry of $\mathbf{A}_{ij}$. At a random $k^{\text{th}}$ time slot, we now can write:
\begin{equation}\label{algorithm13}
\Tilde{p}_f(k+1)=W(k)\Tilde{p}_f(k),
\end{equation}
where the random variable $W(k)$ is drawn i.i.d from some distribution on the set of all possible values $W_{ij}$\cite{boyd2006}. Thus Theorem~\ref{steady} implies that the expectation of the updating matrix ${E}[W(k)]$ is stable, i.e., $-1<\lambda({E}[W(k)])<1$.

To analyze the convergence of the second moment, we obtain the following equation \cite{boyd2006}
\begin{equation}\label{minh}
    \begin{aligned}
        {\rm E}[\Tilde{p}_f(k+1)^{\top}&\Tilde{p}_f(k+1)|\Tilde{p}_f(k)] \\
        &= \Tilde{p}_f(k)^{\top}{\rm E}[W(k)^{\top}W(k)]\Tilde{p}_f(k).
    \end{aligned}
\end{equation}

 It is easy to see that $W(k)^\top W(k)$ is also a random variable which is drawn i.i.d from some distribution on the set of possible values $W_{ij}^\top W_{ij}$ (with a probability $\frac{1}{n}{\rm P}_{ij}$). 

 \begin{theorem}\label{2mm}
 Selecting $\alpha$ such that $\alpha<\text{min}(\frac{2}{\lambda_{\max}(\mathbf{L}^{\rm M}_{ff})},\frac{2}{{\max}||\mathbf{A}_{ij}||})$, under Assumption~\ref{bearing rigid} and $n_a \geq 2$, the spectral radius of ${\rm E}[W(k)^\top W(k)]$ is strictly less than 1, which implies that the proposed algorithm's second moment converges as $k\to \infty$.
\end{theorem}

\begin{IEEEproof} By choosing the updating step sizes $\alpha$ to satisfy Theorem~\ref{2mm}, it can be obtained that each possible $W_{ij}$ has eigenvalues that satisfy $-1<\lambda(W_{ij})\leq 1$ and thus $0\leq\lambda(W_{ij}^\top W_{ij})\leq 1$. Denote $\left\{v_{ij}\right\}$ as the eigenspace of $W_{ij}$ corresponding to the eigenvalue $\lambda=1$. Clearly, $\left\{v_{ij}\right\}$ is also the eigenspace corresponding to the unity eigenvalue of $W_{ij}^\top W_{ij}$. We now treat the expectation ${\rm E}[W_{ij}]$ (resp., ${\rm E}[W_{ij}^\top W_{ij}]$)  as a convex combination of all possible $W_{ij}$ (resp., $W_{ij}^\top W_{ij}$) where ${\rm P}_{ij}\neq 0$. Because $W_{ij}$ is symmetric (and thus $W_{ij}^\top W_{ij}$), ${\rm E}[W_{ij}]$ cannot have a unity eigenvalue unless there exists a common eigenvector between every eigenspace $\left\{v_{ij}\right\}$. From Theorem \ref{steady}, we already have $-1<\lambda({\rm E}[W_{ij}])<1$, which implies $\underset{{\rm P}_{ij}\neq 0}{\bigcap}\left\{v_{ij}\right\}=~\varnothing$. Thus, it is obvious that $0\leq \lambda({\rm E}[W_{ij}^\top W_{ij}])<1$. This completes the proof.\end{IEEEproof}

\subsection{Convergence Rate} 
Inspired by \cite{boyd2006,arxivminh}, we first introduce a quantity of interest
\begin{define}
 \emph{($\epsilon$-convergence time)} For any $0<\epsilon<1$, the $\epsilon$-consensus time is defined as follows:
\begin{equation}
\begin{aligned}
T(\epsilon)=\underset{{\hat{p}_f}(0)}{\sup}\inf \bigg( k:{\rm P}\bigg(\frac{\lVert \hat{p}_f(k)- {p}_f\rVert}{\lVert {\hat{p}_f}(0)-p_f\rVert} \geq \epsilon\bigg)\leq \epsilon\bigg).
\end{aligned}
\end{equation}
\end{define}

Intuitively, $T({\epsilon})$ represents the number of clock ticks needed for the estimator $\hat{p}_f(k)$ to be close to the actual position $p_f$ with a high probability. In this paper, we provide the upper bound formula for the proposed network localization algorithm. 

Next, we have the following derivation according to Theorem \ref{2mm}:
\begin{align*} \label{largesteig}
\Tilde{p}_f(k)^\top &{\rm E}[W(k)^\top W(k)] \Tilde{p}_f(k) \nonumber\\
      &\leq \lambda_{\max}({\rm E}[W(k)^\top W(k)]) \Tilde{p}_f(k)^\top \Tilde{p}_f(k)\\
      &\leq \lambda^k_{\max}({\rm E}[W(k)^\top W(k)]) \Tilde{p}_f(0)^\top \Tilde{p}_f(0).
\end{align*}

We can now state the main result of this subsection in the following theorem.
\begin{theorem}
    \label{upperbound2}
     Under Assumption~\ref{bearing rigid}, if the network has at least two beacons, by selecting a common step size to satisfy Theorem \ref{2mm}, the estimation $\hat{p}_f(k)$ converges in expectation to the actual position $p_f$. Furthermore, the $\epsilon$-convergence time is upper bounded by a function of the spectral radius of ${\rm E}[W(k)^\top W(k)])$.
\end{theorem}
\begin{IEEEproof} Using the Markov's inequality (Lemma 2), we have
\begin{align*}
        {\rm P}\bigg(\frac{\lVert \hat{p}_f(k)-p_f \rVert}{\lVert \hat{p}_f(0)-p_f \rVert}  \geq \epsilon \bigg) 
        &={\rm P}\bigg(\frac{\Tilde{p}_f(k)^\top \Tilde{p}_f(k)}{\Tilde{p}_f(0)^\top \Tilde{p}_f(0)}
        \geq \epsilon^2\bigg) \nonumber\\
        &\leq\frac{\epsilon^{-2}{\rm E}[\Tilde{p}_f(k)^\top \Tilde{p}_f(k)]}{\Tilde{p}_f(0)^\top \Tilde{p}_f(0)}\\
        &\leq \epsilon^{-2} \lambda_{\max}^k \big( {\rm E}[{W}(k)^\top {W}(k)] \big).
\end{align*}
As a result, for $k \geq K(\epsilon)=\frac{3{\log}(\epsilon^{-1})}{{\log}\lambda_{max}^{-1} \big( {\rm E}[{W}(k)^\top {W}(k)] \big)}$, there holds
$${\rm P}\bigg(\frac{\lVert \hat{p}_f(k)- {p_f}\rVert}{\lVert {\hat{p}_f}(0)-p_f\rVert} \geq \epsilon\bigg)\leq \epsilon.$$ 
Thus, $K(\epsilon)$ is the upper bound of the $\epsilon$-consensus time. \end{IEEEproof}

\section{SIMULATION EXAMPLE}\label{simulation}
\begin{figure*}[ht]
    \centering
    \subfloat[The graph $\mathcal{G}$.]{\includegraphics[width=0.3\textwidth]{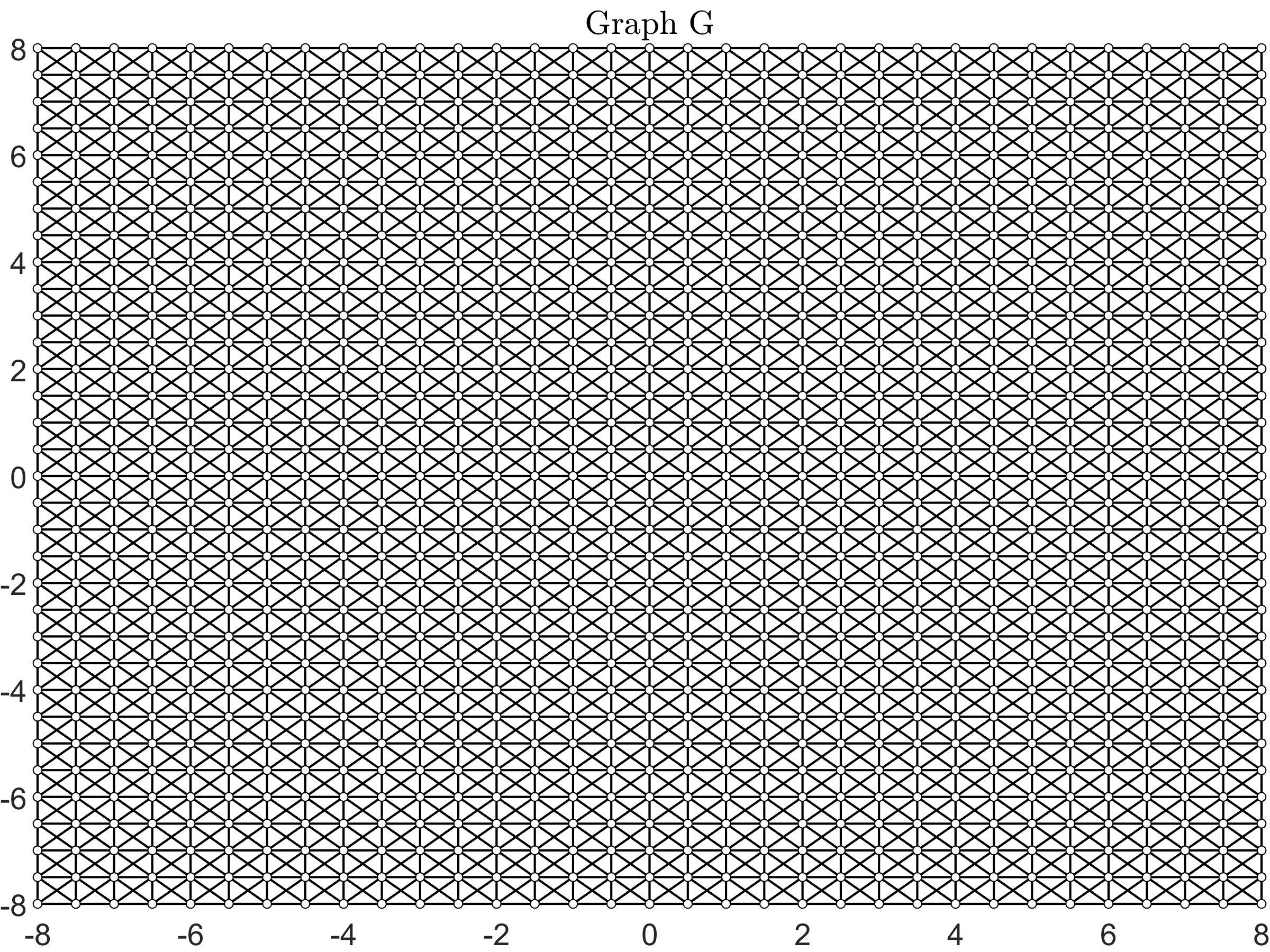} \label{graphG}} \hfill
    \subfloat[The actual positions of sensors.]{\includegraphics[width=0.3\textwidth]{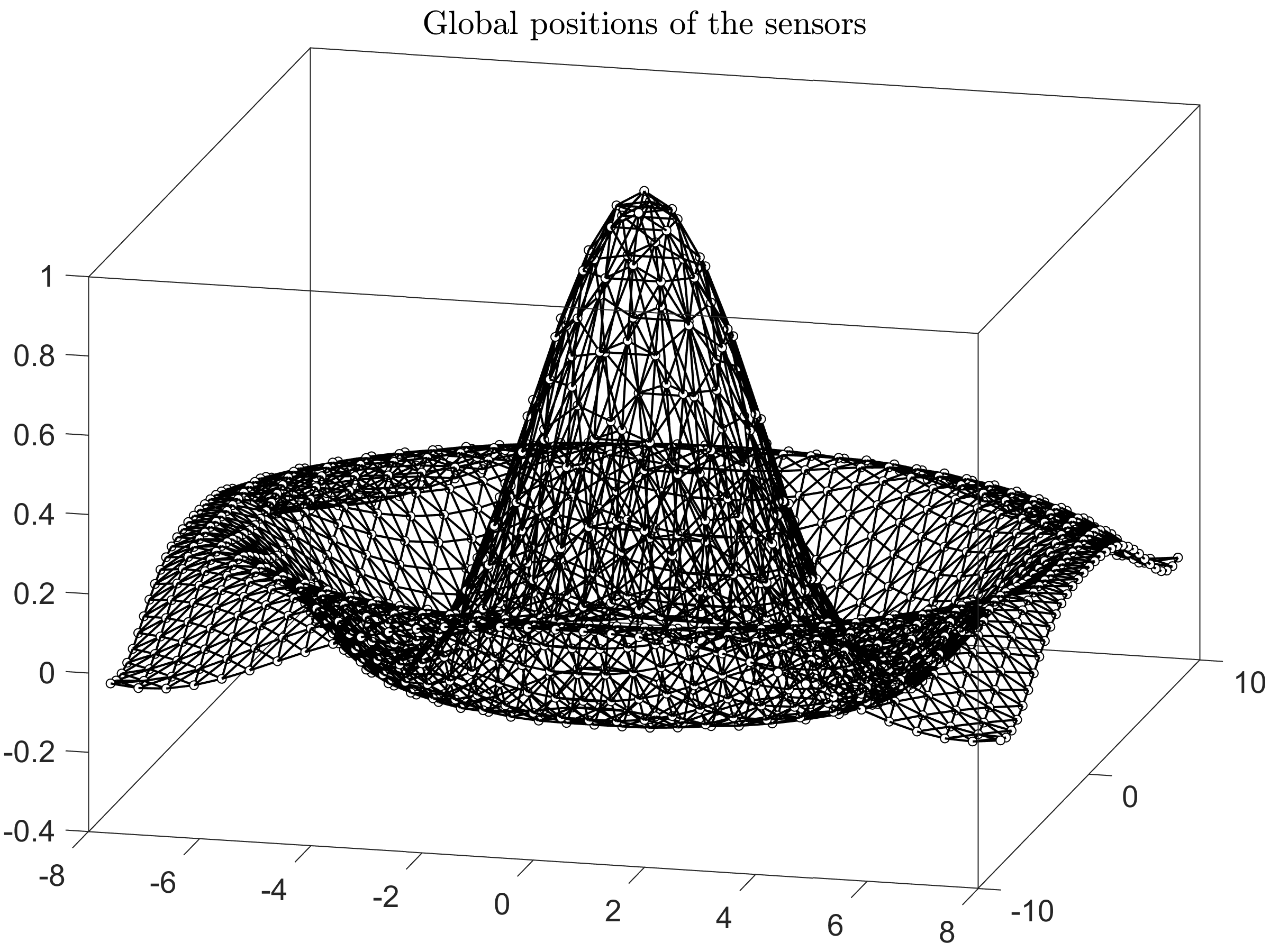}\label{global position}}\hfill
    \subfloat[Total bearing error through time.]{\includegraphics[width=0.3\textwidth]{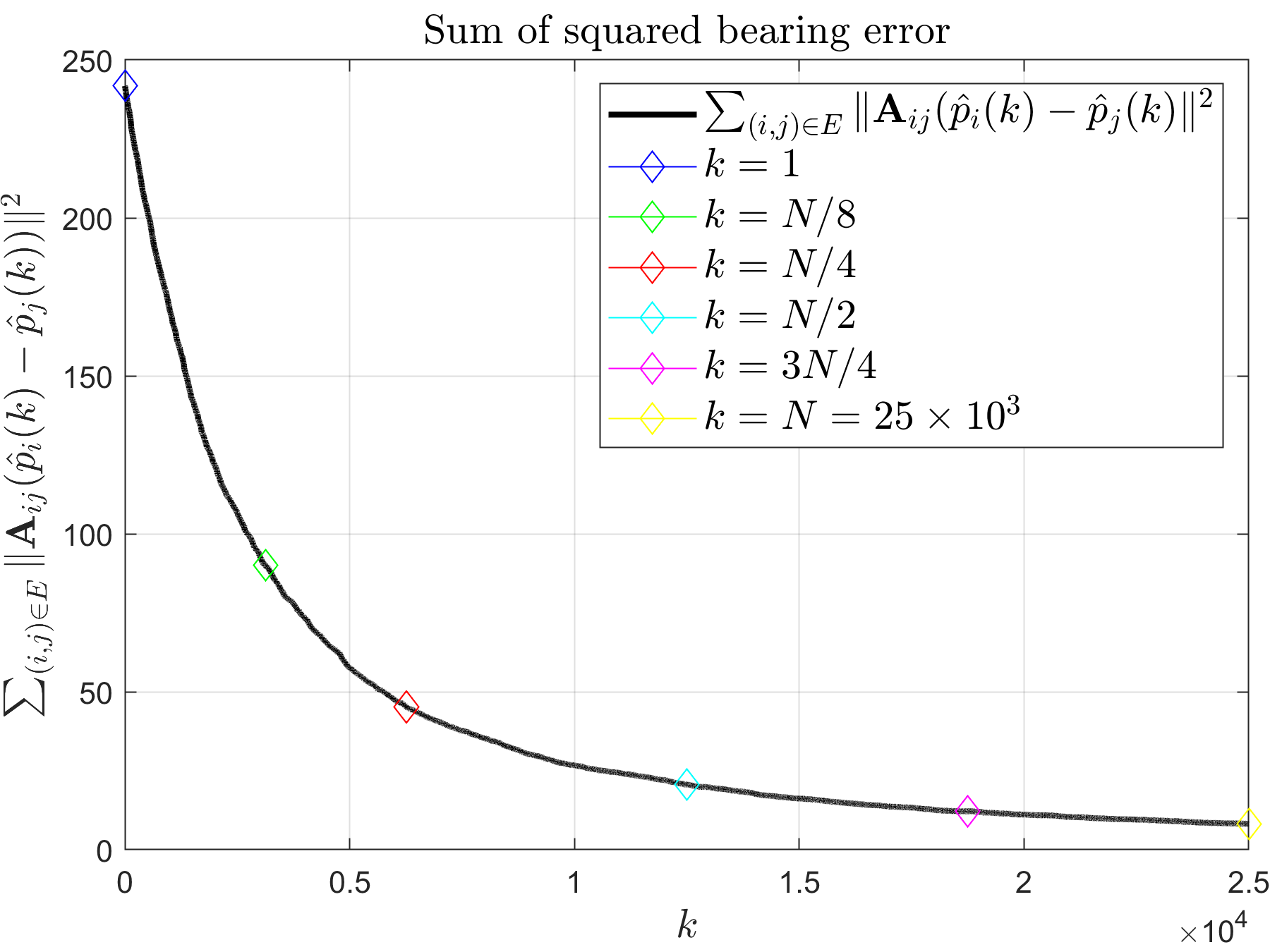}\label{Berror}}\\
    \subfloat[$\hat{p}(k)$ at $k=0$.]{\includegraphics[width=0.3\textwidth]{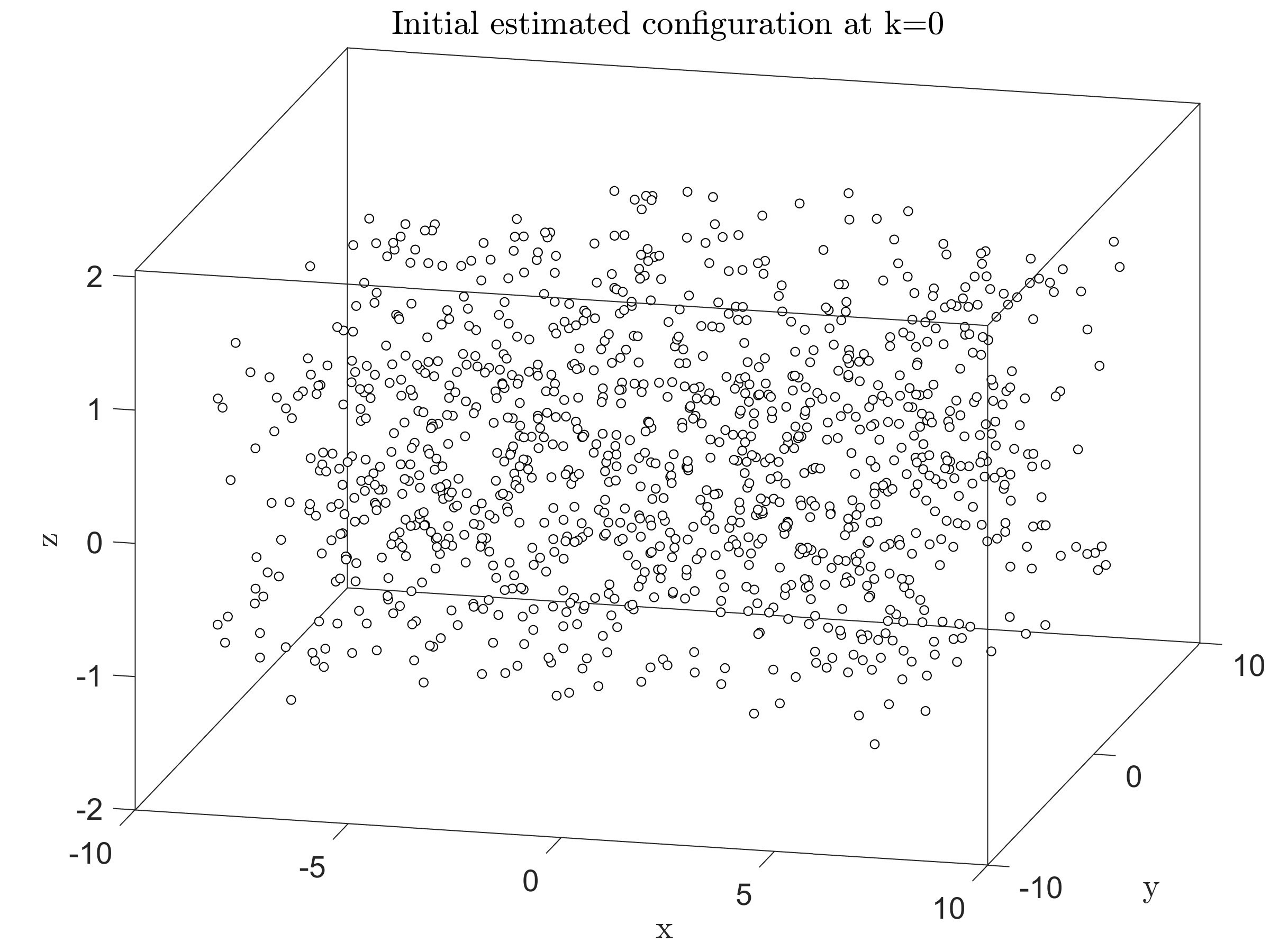}\label{est0}} \hfill
    \subfloat[$\hat{p}(k)$ at $k=N/8$.]{\includegraphics[width=0.3\textwidth]{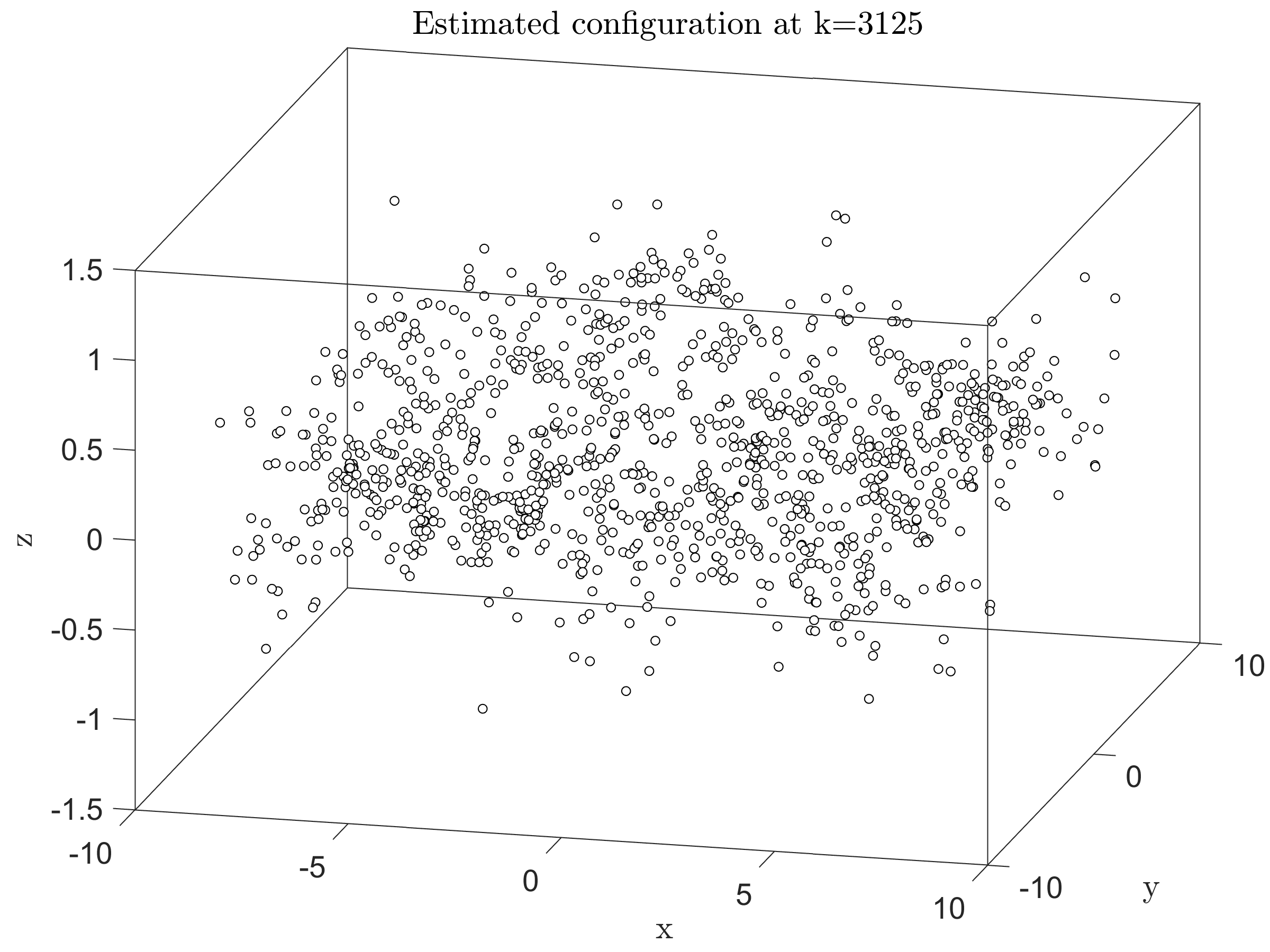}\label{est1}} \hfill
    \subfloat[$\hat{p}(k)$ at $k=N/4$.]{\includegraphics[width=0.3\textwidth]{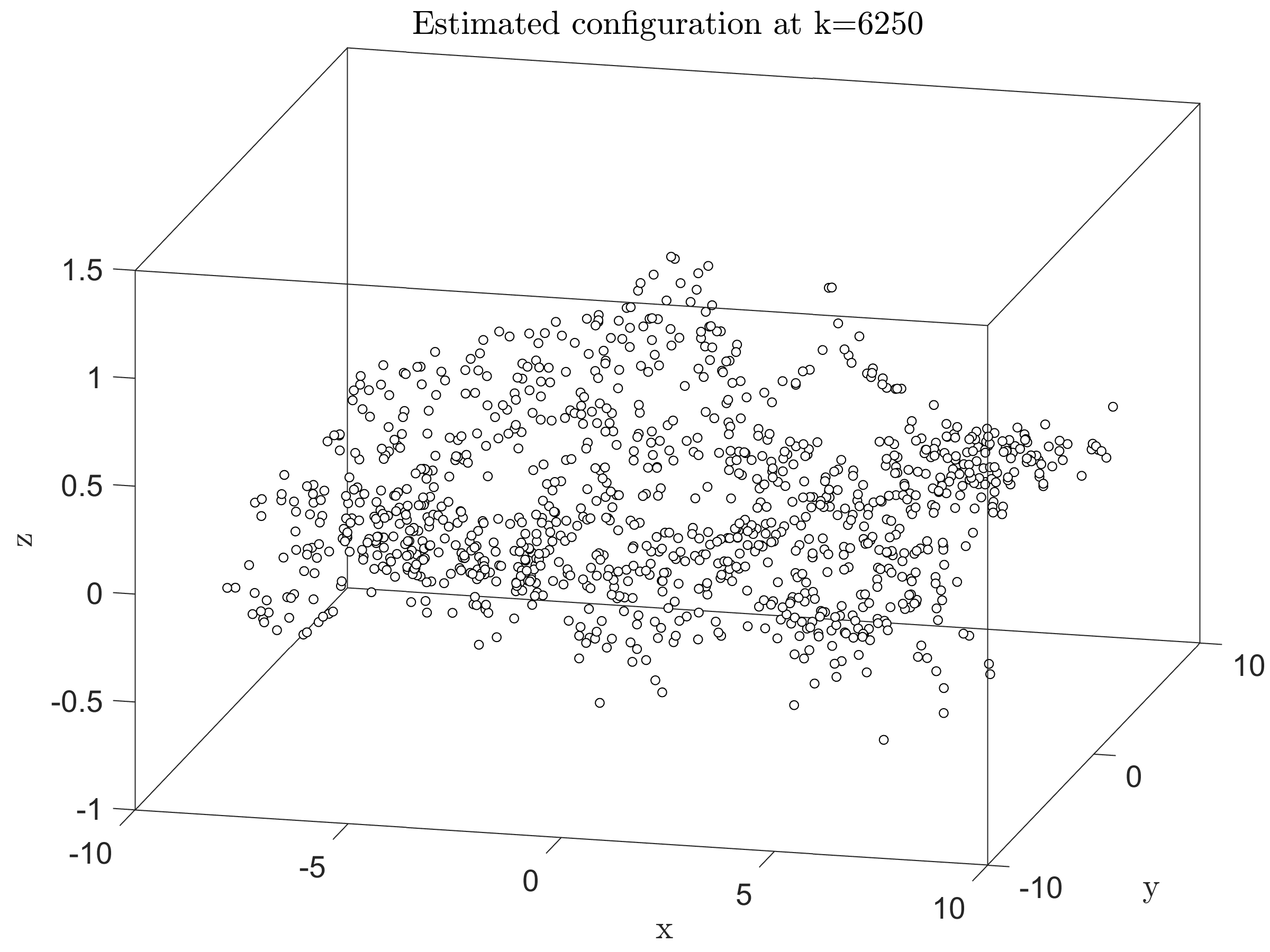}\label{est2}}\\
    \subfloat[$\hat{p}(k)$ at $k=N/2$.]{\includegraphics[width=0.3\textwidth]{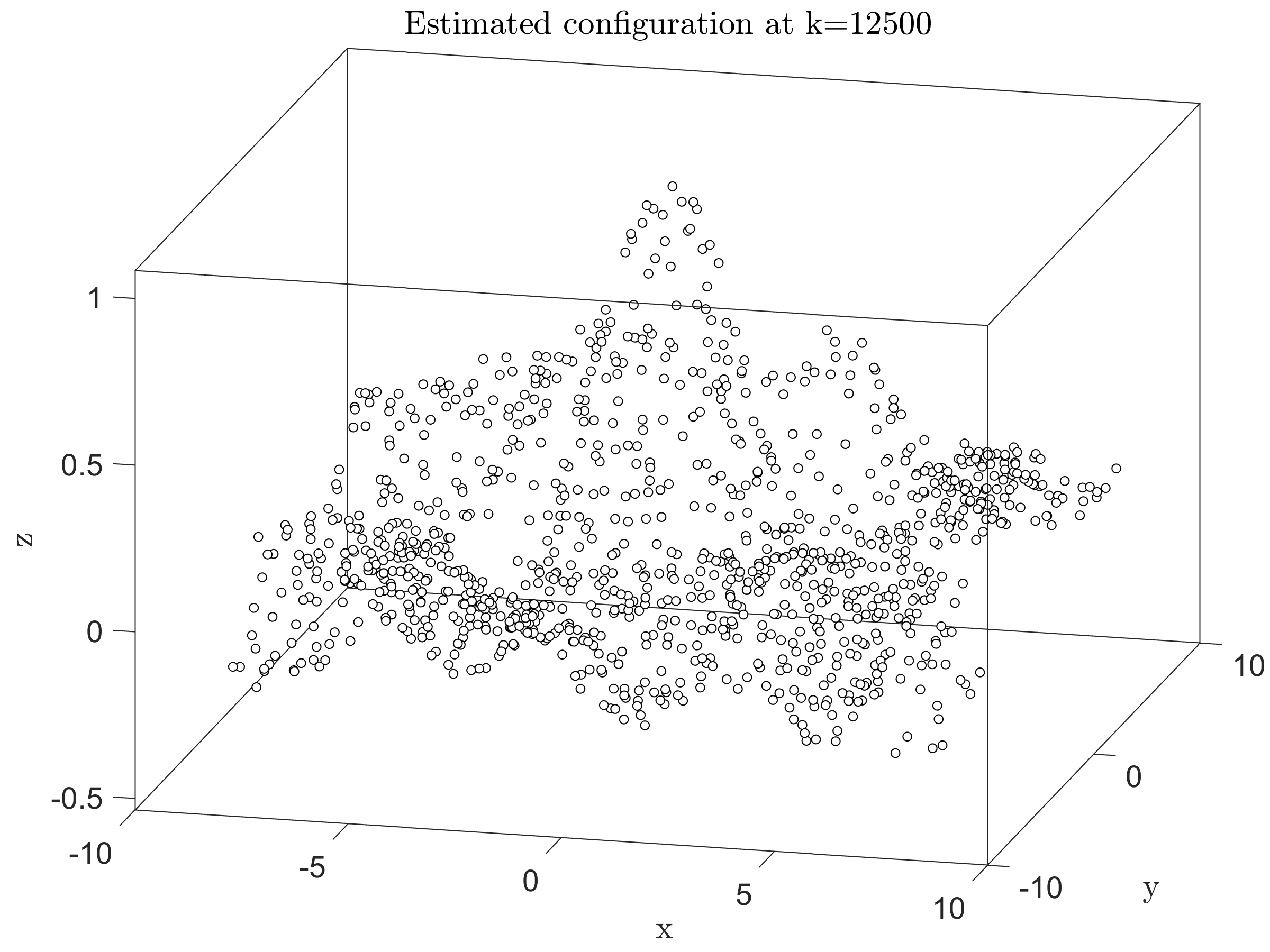}\label{est3}} \hfill
    \subfloat[$\hat{p}(k)$ at $k=3N/4$.]{\includegraphics[width=0.3\textwidth]{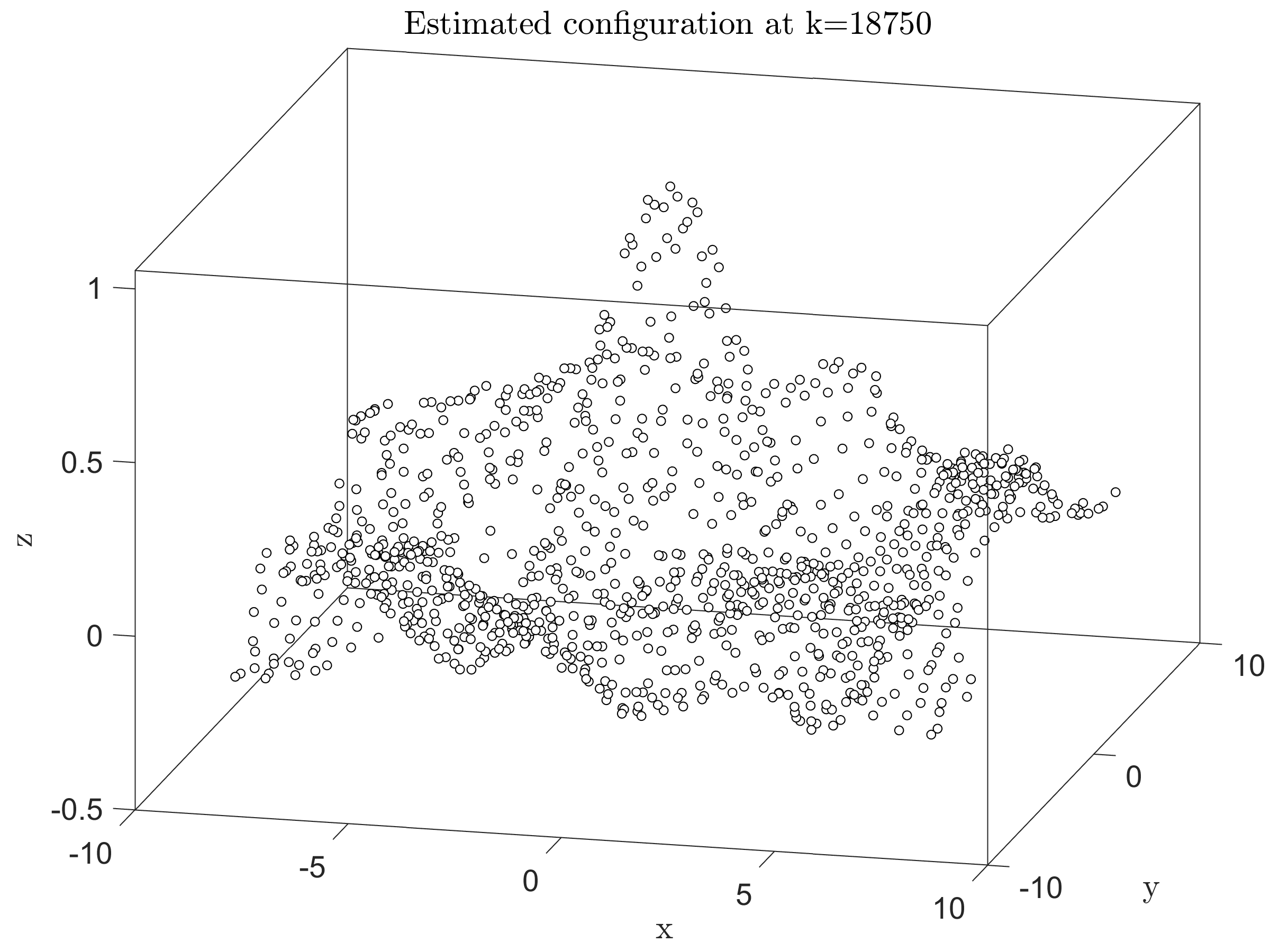}\label{est4}} \hfill
    \subfloat[$\hat{p}(k)$ at $k=N=25\times 10^3$.]{\includegraphics[width=0.3\textwidth]{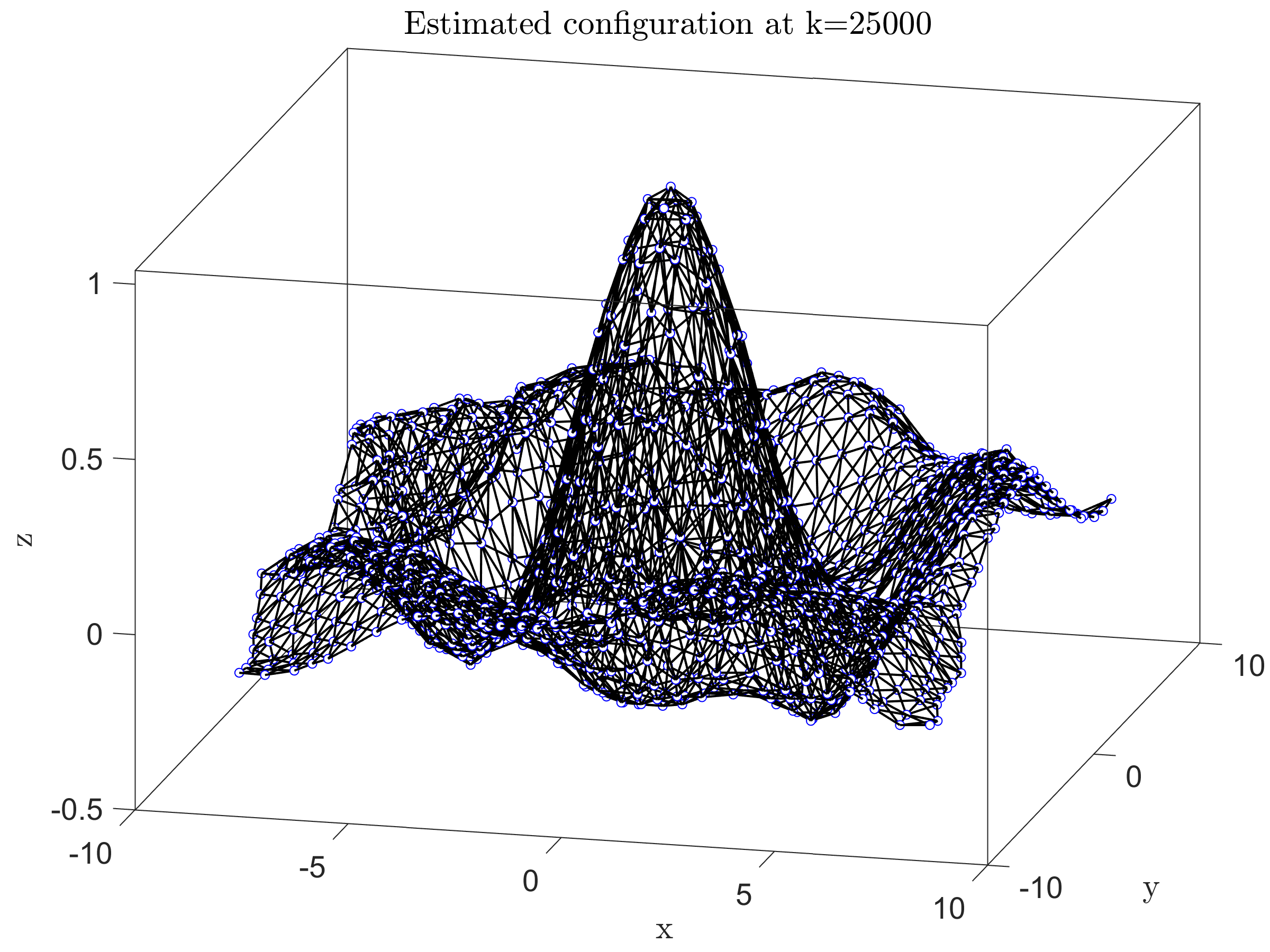}\label{est5}}
    \caption{Simulation of a sensor network consisting of 1089 nodes under the gossip-based network localization protocol \eqref{algorithm0}, \eqref{algorithm1}, \eqref{algorithm2}: (a) - the graph $\mathcal{G}$; (b) - the actual configuration $p$; (c) - the bearing error vs time; From (d) to (i) - the estimate configurations at different time instances.}
\end{figure*}

 Consider a network of $n=1089$ sensor nodes in a three-dimensional space ($d=3$), with $p_i=[x_i,y_i,z_i]^{\top}$. There are $n_a=2$ beacons (nodes 1 and 2) in the network. As depicted in Fig.~\ref{global position}, the $x-$ and $y-$ coordinates of the sensors are distributed evenly along an $x-$, $y-$ mesh given by $x=[-8:0.5:8],~
 y=[-8:0.5:8].$ Meanwhile, the $z-$coordinates of $n$ sensors satisfy
$$z=\frac{\text{sin}(\sqrt{x^2+y^2})}{\sqrt{x^2+y^2}}.$$
The initial estimation $\hat{p}_i(0)=[\hat{x}_i(0),\hat{y}_i(0),\hat{z}_i(0)]^\top$ of each follower node is generated randomly in a cubic $[-8,8]\times[-8,8]\times[-8,2]$, which is shown in Fig. \ref{global position}. The edges in $E$, being chosen accordingly to the proximity-rule
\[(i,j) \in E \Longleftrightarrow \|p_i - p_j\|\leq \frac{\sqrt{2}}{2},\]
results to the topological graph $G$ in Figure \ref{graphG}.

The simulation result of the sensor network under the randomized network localization protocol \eqref{algorithm0}, \eqref{algorithm1}, \eqref{algorithm2} is illustrated in Fig.~\ref{Berror}-\ref{est5}. As can be shown in Fig.~\ref{est0}--\ref{est5}, snapshots of the estimate configuration $\hat{p}(k)$ at time instances $k =0, N/8, N/4, N/2, 3N/4, N$, for $N=25\times 10^3$, demonstrate that all position estimates eventually converge to the true values as $k\to \infty$. Additionally, it can be seen from Fig.~\ref{Berror} that the total bearing error, which is defined as $\sum_{(i,j)\in E} \|\mathbf{A}_{ij}(\hat{p}_j(k)-\hat{p}_i(k))\|^2$, converges to 0 over time at exponential rate.

Thus, the simulation result is consistent with the convergence analysis.

% \begin{figure}[!t]
% \centerline{\includegraphics[width=\columnwidth]{global position.png}}
% \caption{The actual positions of sensors.}
% \label{global position}
% \end{figure}

% \begin{figure}[!t]
% \centerline{}
% \caption{}

% \end{figure}
% \begin{figure}[!t]
% \centerline{}
% \caption{}

% \end{figure}

% \begin{figure}[!t]
% \centerline{}
% \caption{}

% \end{figure}\textbf{}

\section{Conclusion}\label{conclusion}
In this paper, we propose a bearing-based network localization algorithm under the gossip protocol to estimate the positions of nodes in a wireless sensor network. The convergence of expectation and second moment of estimation errors were rigorously proven. The theoretical result is confirmed by the numerical example. A drawback of the algorithm is that the upper bound of the update step-size is dependent on the maximum eigenvalue of the grounded Laplacian $\mathbf{L}_{ff}^{\rm M}$, which is usually a quantity that can only be estimated by the agents. A future research direction is to improve the convergence speed of the algorithm. It is also interesting to extent the algorithm so that more than two neighboring agents can update their estimates at the same time slot.
%\appendices

%\section*{Acknowledgment}

%\section*{References and Footnotes}

%\subsection{References}

\bibliographystyle{IEEEtran}
\bibliography{ref}

% Generated by IEEEtran.bst, version: 1.14 (2015/08/26)
\begin{thebibliography}{10}
\providecommand{\url}[1]{#1}
\csname url@samestyle\endcsname
\providecommand{\newblock}{\relax}
\providecommand{\bibinfo}[2]{#2}
\providecommand{\BIBentrySTDinterwordspacing}{\spaceskip=0pt\relax}
\providecommand{\BIBentryALTinterwordstretchfactor}{4}
\providecommand{\BIBentryALTinterwordspacing}{\spaceskip=\fontdimen2\font plus
\BIBentryALTinterwordstretchfactor\fontdimen3\font minus
  \fontdimen4\font\relax}
\providecommand{\BIBforeignlanguage}[2]{{%
\expandafter\ifx\csname l@#1\endcsname\relax
\typeout{** WARNING: IEEEtran.bst: No hyphenation pattern has been}%
\typeout{** loaded for the language `#1'. Using the pattern for}%
\typeout{** the default language instead.}%
\else
\language=\csname l@#1\endcsname
\fi
#2}}
\providecommand{\BIBdecl}{\relax}
\BIBdecl

\bibitem{Aspnes2006}
J.~Aspnes, T.~Eren, D.~K. Goldenberg, A.~Morse, W.~Whiteley, Y.~R. Yang,
  B.~D.~O. Anderson, and P.~Belhumeur, ``A theory of network localization,''
  \emph{IEEE Transactions on Mobile Computing}, vol.~5, no.~12, pp. 1663--1678,
  2006.

\bibitem{MAO20072529}
G.~Mao, B.~Fidan, and B.~D.~O. Anderson, ``Wireless sensor network localization
  techniques,'' \emph{Computer Networks}, vol.~51, no.~10, pp. 2529--2553,
  2007.

\bibitem{Ye2017bearing}
M.~Ye, B.~D.~O. Anderson, and C.~Yu, ``Bearing-only measurement
  self-localization, velocity consensus and formation control,'' \emph{IEEE
  Transactions on Aerospace and Electronic Systems}, vol.~53, no.~2, pp.
  575--586, 2017.

\bibitem{Bishop2009bearing}
A.~N. Bishop, B.~D.~O. Anderson, B.~Fidan, P.~N. Pathirana, and G.~Mao,
  ``Bearing-only localization using geometrically constrained optimization,''
  \emph{IEEE Transactions on Aerospace and Electronic Systems}, vol.~45, no.~1,
  pp. 308--320, 2009.

\bibitem{zhu2014}
G.~Zhu and J.~Hu, ``A distributed continuous-time algorithm for network
  localization using angle-of-arrival information,'' \emph{Automatica},
  vol.~50, p. 53–63, 01 2014.

\bibitem{Zhong2014}
J.~Zhong, Z.~Lin, Z.~Chen, and W.~Xu, ``Cooperative localization using
  angle-of-arrival information,'' \emph{Proc. of the 11th IEEE International
  Conference on Control \& Automation (ICCA)}, pp. 19--24, 2014.

\bibitem{ZHAO2016334}
Z.~Shiyu and D.~Zelazo, ``Localizability and distributed protocols for
  bearing-based network localization in arbitrary dimensions,''
  \emph{Automatica}, vol.~69, pp. 334--341, 2016.

\bibitem{Li2020tcns}
X.~Li, X.~Luo, and S.~Zhao, ``Globally convergent distributed network
  localization using locally measured bearings,'' \emph{IEEE Transactions on
  Control of Network Systems}, vol.~7, no.~1, pp. 245--253, 2020.

\bibitem{Cao2021icsmc}
M.~Cao, H.~Zhang, Z.~Wang, C.~Zhang, and C.~Huang, ``Fixed-time bearing-based
  distributed network localization,'' in \emph{2021 IEEE International
  Conference on Systems, Man, and Cybernetics (SMC)}, 2021, pp. 964--969.

\bibitem{boyd2006}
S.~Boyd, A.~Ghosh, B.~Prabhakar, and D.~Shah, ``Randomized gossip algorithms,''
  \emph{IEEE Transactions on Information Theory}, vol.~52, no.~6, pp.
  2508--2530, 2006.

\bibitem{ishii2010}
H.~Ishii and R.~Tempo, ``Distributed randomized algorithms for the pagerank
  computation,'' \emph{IEEE Transactions on Automatic Control}, vol.~55, no.~9,
  pp. 1987--2002, 2010.

\bibitem{STRAKOVA2013IOP}
H.~Strakov{\'a}, G.~Niederbrucker, and W.~N. Gansterer, ``Fault tolerance
  properties of gossip-based distributed orthogonal iteration methods,''
  \emph{Procedia Computer Science}, vol.~18, pp. 189--198, 2013.

\bibitem{5625615}
W.~Li, H.~Dai, and Y.~Zhang, ``Location-aided fast distributed consensus in
  wireless networks,'' \emph{IEEE Transactions on Information Theory}, vol.~56,
  no.~12, pp. 6208--6227, 2010.

\bibitem{TIT2010}
F.~Bénézit, A.~G. Dimakis, P.~Thiran, and M.~Vetterli, ``Order-optimal
  consensus through randomized path averaging,'' \emph{IEEE Transactions on
  Information Theory}, vol.~56, no.~10, pp. 5150--5167, 2010.

\bibitem{1238221}
D.~Kempe, A.~Dobra, and J.~Gehrke, ``Gossip-based computation of aggregate
  information,'' in \emph{44th Annual IEEE Symposium on Foundations of Computer
  Science, 2003. Proceedings.}, 2003, pp. 482--491.

\bibitem{7581065}
A.~Khosravi and Y.~S. Kavian, ``Broadcast gossip ratio consensus: Asynchronous
  distributed averaging in strongly connected networks,'' \emph{IEEE
  Transactions on Signal Processing}, vol.~65, no.~1, pp. 119--129, 2017.

\bibitem{4787122}
T.~C. Aysal, M.~E. Yildiz, A.~D. Sarwate, and A.~Scaglione, ``Broadcast gossip
  algorithms for consensus,'' \emph{IEEE Transactions on Signal Processing},
  vol.~57, no.~7, pp. 2748--2761, 2009.

\bibitem{8946047}
X.~He, Y.~Cui, and Y.~Jiang, ``An improved gossip algorithm based on
  semi-distributed blockchain network,'' in \emph{Proc. of the International
  Conference on Cyber-Enabled Distributed Computing and Knowledge Discovery
  (CyberC)}, 2019, pp. 24--27.

\bibitem{Liu2013analysis}
J.~Liu, B.~D.~O. Anderson, M.~Cao, and A.~S. Morse, ``Analysis of accelerated
  gossip algorithms,'' \emph{Automatica}, vol.~49, no.~4, pp. 873--883, 2013.

\bibitem{arxivminh}
N.-M. Le-Phan, M.~H. Trinh, and P.~D. Nguyen, ``Randomized matrix-weighted
  consensus,'' \emph{arxiv preprint: https://arxiv.org/abs/2303.14733}, 2023.

\bibitem{HMT2018}
M.~H. Trinh, C.~V. Nguyen, Y.-H. Lim, and H.-S. Ahn, ``Matrix-weighted
  consensus and its applications,'' \emph{Automatica}, vol.~89, 01 2018.

\bibitem{zhaoTAC}
Z.~Shiyu and D.~Zelazo, ``Bearing rigidity and almost global bearing-only
  formation stabilization,'' \emph{IEEE Transactions on Automatic Control},
  vol.~61, no.~5, pp. 1255--1268, 2016.

\bibitem{prob}
D.~Bertsekas and J.~Tsitsiklis, \emph{Introduction to probability}.\hskip 1em
  plus 0.5em minus 0.4em\relax Massachusetts: Athena Scientific, 2008.

\bibitem{bearingzhao}
S.~Zhao and D.~Zelazo, ``Bearing-based distributed control and estimation of
  multi-agent systems,'' in \emph{Proc. of the European Control Conference
  (ECC), Linz, Austria}, 2015, pp. 2202--2207.

\end{thebibliography}
\end{document}